# Matching the Spectral Energy Distribution and p Mode Oscillation Frequencies of the Rapidly Rotating Delta Scuti Star α Ophiuchi with a 2D Rotating Stellar Model

Robert G. Deupree, Diego Castañeda, Fernando Peña, and C. Ian Short

Institute for Computational Astrophysics and Department of Astronomy and Physics,

Saint Mary's University, Halifax, NS  B3H 3C3 Canada; bdeupree@ap.smu.ca

ABSTRACT

Spectral energy distributions are computed using 2D rotating stellar models and NLTE plane parallel model atmospheres. A rotating, 2D stellar model has been found which matches the observed ultraviolet and visible spectrum of α Oph. The SED match occurs for the interferometrically deduced surface shape and inclination, and is different from the SED produced by spherical models. The p mode oscillation frequencies in which the latitudinal variation is modelled by a linear combination of eight Legendre polynomials were computed for this model. The five highest and seven of the nine highest amplitude modes show agreement between computed axisymmetric, equatorially symmetric mode frequencies and the mode frequencies observed by MOST to within the observational error. Including nonaxisymmetric modes up through $|m| = 2$ and allowing the possibility that the eight lowest amplitude modes could be produced by modes which are not equatorially symmetric produces matches for 24 out of the 35 MOST modes to within the observational error and another eight modes to within twice the observational error. The remaining three observed modes can be fit within 4.2 times the observational error, but even these may be fit to within the observational error if the criteria for computed modes are expanded.

## 1. INTRODUCTION

While much progress has been made in understanding the structure and evolution of stars over the last 50 years, most of the gains have been for spherical stars. Models which contain a significant nonspherical component, such as rotating stars, are much less well understood. Part of this must be put down to the increased degrees of freedom rotation introduces, both into the models (angular momemtum distribution, possible mixing) and into the observations (the effects of the latitudinal surface variations and the inclination between the observer and the rotation axis in converting the observed effective temperature and luminosity into properties physically associated with the star). It has been hoped that the combination of asteroseismology and interferometry would allow more progress in the understanding of rotating stars (Cunha et al. 2007), but to date progress has been relatively slow. An early attempt to determine the shape of the rapidly rotating star Achenar (Domiciano de Souza, et al. 2003) was clouded by the possibility that a circumstellar envelope was contributing to the oblateness measured by the interferometry as well as the stellar surface (Vinicius, et al., 2006; Kanaan, et al. 2008; Carcofi, et al. 2008; Kervella, et al. 2009), but it did lead to some attempts which successfully reproduced the observed shape (Jackson, et al. 2004).

A more promising candidate, α Oph, has recently surfaced. It has been observed interferometrically with the CHARA array (Zhao et al. 2009) and asteroseismologically with MOST (Monnier et al. 2010). It also helps that the star is so close that it has a well determined parallax (e.g., Gatewood 2005) and virtually no reddening. We summarize the key observational parameters in Table 1**,** the mass coming from a recent analysis of the binary system of which the star of interest in this work is the primary (Hinkley, et al. 2011).

We wish to make 2D rotating models of α Oph. With these models and some of the data from Table 1(listed as "assumed") we will compute the other data from Table 1, the spectral energy distribution (SED) from the ultraviolet through the visible, and the oscillation frequencies for comparison with their observed counterparts. The computed surface of the 2D finite difference model is discretized (i.e., a zone boundary of the 2D mesh is used to define the surface location at each latitude), so the observed and assumed ratio of the polar to the equatorial surface radius are not exactly the same. However, the assumed values are within the error of the observed values both for the polar to equatorial ratio and for the mass.

The computation of multi-dimensional structural information for (at least conservatively) rotating models with reasonable physical input has been possible for some time (e.g., Clement 1978, 1979; Deupree 1990, 1995, 2011a; Jackson, et al. 2005; Espinosa Lara 2010). This usually involves imposing a composition profile and a rotation profile and then solving for the model structure. However, performing multi-dimensional simulations of rotating stars in which the angular momentum and composition are redistributed during evolution remains incomplete (e.g., Tassoul & Tassoul 1982, 1995; Espinosa Lara & Rieutord 2007; Rieutord & Espinosa Lara 2009). The need to compute oscillation frequencies by methods which allow the latitudinal variation of the eigenfunctions to be determined by a sum of spherical harmonics was first outlined by Berthomieu, et al. (1978) with pioneering work developed by Lee & Saio (1986), Lee & Baraffe (1996), Aprilia et al. (2011). This work has been extended (e.g., Clement 1998; Lignières et al. 2006; Reese et al. 2006, 2008; Lovekin et al. 2009;) to include realistic input models of rotating stars (Lovekin & Deupree 2008; Reese et al. 2009). These approaches give reasonably reliable results, although gains are still possible. This more complex representation of the latitudinal variation does make mode identification more complicated, as does the fact that the frequency splitting for the nonaxisymmetric modes is both large and nonlinear (e.g., Suárez et al. 2010; Deupree & Beslin 2010).

To obtain the SED we follow the work of Lovekin et al. (2006) and Gillich et al. (2008), who used 2D rotating stellar structure models to provide the latitudinal surface temperature and effective gravity variation. This was used with PHOENIX (Hauschildt & Barron 1999) NLTE plane parallel model atmospheres to compute the flux as a weighted sum of the radiative intensities from the surface in the direction of the observer, integrated over the visible surface of the model. This method of determining the observed flux is the same in concept as used by other researchers (e.g., Slettebak et al. 1980; Linnell & Hubeny 1994; Frémat et al. 2005; Dall & Sbordone 2011).

The goal of this paper is to produce a 2D model of α Oph which has the observed oblateness and inclination (both assumed), matches the observed SED when placed at the proper distance, and matches the observed oscillation modes. In the next section we describe how the 2D models were generated and refined to fit the observed SED. We also present some details of the model atmosphere calculations. The third section computes oscillation frequencies for the p modes for the best fit model to the SED. The final section reviews the results and the assumptions made.

## 2. 2D STELLAR MODELS, NLTE MODEL ATMOSPHERES, AND THE OBSERVED SPECTRAL ENERGY DISTRIBUTION

In order to compute the SED of a rotating star we must first know the latitudinal variation of the surface temperature and effective gravity. Starting with the model V240, which has the observed oblateness, generated by Deupree (2011b), we use the 2D stellar evolution and hydrodynamics code ROTORC (Deupree 1990, 1995) to make models which provide this information. This code allows us to determine the full 2D model structure for a rotating stellar model. Model V240 was constructed with uniform rotation, which we retain for all the models in this work. Comparison between the SED of model V240 seen at an inclination of 87.5$^{o}$ and the observed SED in the visual and ultraviolet indicated that this model has too high an effective temperature. New models were obtained by performing a few evolutionary time steps (thus moving to cooler effective temperatures), which change the composition in the convective core and the surrounding area as the convective core shrinks. During this "evolution" we hold the rotational velocity and the surface location constant. The surface rotation velocities are no longer those required for an equipotential surface to match the current (desired) surface shape, so the uniform rotation rate was changed and the model at the end of this evolution sequence reconverged until an equipotential surface matches the desired surface shape. Only a few evolutionary times steps were done at a time so that the change in the surface equatorial velocity was only a few km $s^{-1}$. The entire process is repeated until there are a sufficiently large number of models with composition profiles that look like those at various stages of an evolutionary sequence. From an observational point of view, these models form a sequence of decreasing effective temperature. It should, however, be noted that these are not strictly speaking evolutionary sequences because we have artificially constrained the rotation during the evolution in a way that does not conserve angular momentum. From the point of view of creating a 2D stellar model which can be compared with data, not having obtained the model by a direct evolutionary sequence calculation is irrelevant.

Once we have a 2D model we compute as a function of wavelength the flux one would observe outside the earth's atmosphere for a star at a specified distance. This flux is a weighted integral of the intensity integrated over the visible surface of the star (Lovekin, et al. 2006). The intensities in the direction of the observer are obtained from interpolation through a grid of plane parallel model atmospheres, using the geometry to indicate the direction to the observer from the local vertical.

The model atmospheres are computed with the PHOENIX code (Hauschildt & Baron 1999). In addition to the composition, the input requirements for the plane-parallel atmospheres used here are the effective temperature and the effective gravity. The composition was taken to be X=0.7, Z=0.02. One of the main advantages of PHOENIX is the capability to include a large number of energy levels for several ionization stages of many elements in Non-LTE (NLTE) (Short et al. 1999). The elements and ionization stages included in NLTE are H, He(I-II), Li(I-II), C(I-III), N(I-III), O(I-III), Ne(I), Na(I-II), Mg(I-III), Al(I-IV), Si(I-IV), P(I-IV), S(I-IV), K(I-III), Ca(I-III), Fe(I-IV). The number of energy levels and line transitions for the ionization stages of the elements included are the same as in Table 1 of Gillich, et al. (2008). For those species treated in NLTE, only energy levels connected by transitions for which log (*gf*) is greater than -3 in the PHOENIX line list are included in the NLTE statistical equilibrium equations. All other transitions for that species are calculated with occupation numbers set equal to the Boltzmann distribution value with the excitation temperature equal to the local kinetic temperature, multiplied by the NLTE departure coefficient for the ground state in the next higher ionization stage.

The energy level and bound-bound transition atomic data have been taken from Kurucz (1994) and Kurucz & Bell (1995). The resonance-averaged Opacity Project (Seaton et al. 1994) data of Bautista et al. (1998) have been used for the ground-state photoionization cross sections of Li (I-II), C (I-IV), N (I-VI), O (I-VI), Ne (I), Na (I-VI), Al (I-VI), Si (I-VI), S (I-VI), Ca (I-VII), and Fe (I-VI). For the ground states of all stages of P and Ti and for the excited states of all species, we used the cross-sectional data previously incorporated into PHOENIX from either Reilman & Manson (1979) or from the compilation of Mathisen (1984). The coupling among all bound levels by electronic collisions is calculated using cross sections calculated from the formulae given by Allen (1973). The cross sections of ionizing collisions with electrons are calculated from the formula of Drawin (1961). Further details are provided by Short et al. (1999).

The model atmosphere grid covers the temperature range from 7500K to 11000K in 250K steps. The effective gravity range is 3.0 to 4.666 in 0.333 steps. The wavelength range starts at 600Å with steps of 0.005Å. The step increases in seven jumps, ending with a step 0.08Å. This step size assures that spectral lines are sufficiently sampled. The final wavelength is 20000Å. The Doppler effect is not included in these particular calculations because we then filter both the observed and computed flux spectra with a 50Å wide rolling boxcar filter to smooth out the effect of lines in the SED.

We have fit several spherical uniform effective temperature models and several rotating models to the observed data in both the visible and the ultraviolet. The data in the visible region was obtained from dataset L1985BURN, file 01398 of the HyperLeda database (Paturel et al. 2003) and the ultraviolet data from datasets SWP17411 and LWR05927 in the IUE database maintained by STScI. The observed data is what would be observed at the earth, or above the earth's atmosphere in the case of the ultraviolet data. The best spherical model match to the visual data has an effective temperature of 7875K. This is shown in Figure 1. The nonrotating, plane parallel atmosphere models do not come from an evolution sequence, so we do not have a radius automatically associated with them. We thus just scale the computed flux to match the data in the visible region as closely as possible**.** In particular, we force the plane parallel model flux to match the observed flux at 5750 Å and compare the observed and model fluxes at 4050Å, an interval between 4400Å and 4470Å, and 6950 Å. The average error magnitude is about 1% for the 7785K and 8000K spherical models, with the lower temperature model being slightly better. Models with effective temperatures of 7750K and 8250K have average errors about a factor of 3 to 5 higher depending on the precise measurement. These models also display the change in SED slope with temperature one expects, making them different from the observed slope. The computed flux for the 7875K model to the blue of the Balmer jump and between 1500 - 3000Å is lower than that observed to varying degrees. Comparisons of the SEDs for spherical models having effective temperatures of 7750K and 8000K with the ultraviolet data just to the blue of the Balmer jump are about the same, and all three spherical models have fluxes smaller than that observed. To quantify the comparison in the ultraviolet, we computed the magnitude of the difference between the observed flux and the computed fluxes every ten Å and then calculated the average of this error magnitude for each of the two IUE datasets. The minimum average error magnitude for all the plane parallel atmospheres in both IUE wavelength intervals was for the 7875K effective temperature model, with the 8000K model being more than twice this minimum and other effective temperature models, both higher and lower, having appreciably higher average error magnitudes. We therefore restrict further discussion related to the SEDs to the 7785K effective temperature model. We also note that the computed fluxes in the core of the strong hydrogen lines are appreciably deeper than those observed for all models, both rotating and not.

The rotating model which best fits the flux an observer at the observed inclination would see has the properties $M = 2.25M_{\odot}$, $X_c = 0.25$, $R_{eq} = 3.006R_{\odot}$, $R_{pole} = 0.838R_{eq}$, $V_{eq} = 236$ km $s^{-1}$,$T_{eq} = 7735K$, and $T_{pole} = 9135K$. The computed flux for this model in the direction of the observed inclination is compared with the observed flux in Figure 2. The first point to make is that the computed flux here is not arbitrarily scaled but generated assuming a distance to α Oph of 14.6 pc, close to the most recent determination and within the error estimate (Gatewood 2005). The luminosity of the model is $33.45L_{\odot}$, while the "observed luminosity" is $23.1L_{\odot}$. This observed luminosity is obtained by integrating this computed flux over all wavelengths and multiplying the result by $4\pi d^2$, where d is the distance to the star. The integration over wavelength includes the assumption of a Rayleigh-Jeans tail from the end of the calculated wavelengths to infinite wavelength. We have computed the colors for some of the Johnson (1954) filters and present the results in Table 1. We performed a single point calibration with a NLTE PHOENIX model of Vega using the parameters of Castelli & Kurucz (1994) as one of the steps to determine the V magnitude and color indices listed in Table 1 for the rotating model of α Oph.

Generally speaking, the fit to the SED for this rotating model, shown in Figure 2, is overall better than that for the spherical models. Specifically, the fit to the data in the visible is about the same as that of the 7875K and 8000K effective temperature models (1%)**,** the fit to the blue of the Balmer jump is better, although still not perfect, and the fit between 2000 - 3000Å is much better than any of the spherical models show. The average error magnitude for this rotating model was the same as for the 7875K model in the far ultraviolet dataset, although comparison of Figures 1 and 2 shows that neither fits the SED very well, but in different ways, over the whole region. We note that the far ultraviolet dataset does not have the prominent peak at about 1600Å that all spherical models and the rotating model possess. This plays a role in the far ultraviolet error comparison, making no model particularly good in at least part of this wavelength region.

The rotating model SED agrees much better than that of the 7875K spherical model for the other IUE ultraviolet dataset. The average error magnitude of the rotating model is only 54% of that for the 7875K spherical model. This better agreement with the observations for the rotating model SED provides some validation both for the rotating models themselves and to the numerical approach of integrating the localized surface temperatures and gravities over the surface to obtain the observed flux.

## 3. OSCILLATION MODE SPECTRUM

We have computed the oscillation mode spectrum for the best fit rotating model found by matching the SED in the previous section. We calculate the linear adiabatic oscillation frequencies with the NRO code (Clement 1998) using a sum of eight spherical harmonics (hereinafter, basis functions) to describe the latitudinal variation of the linear perturbations. We believe this will provide reasonable accuracies for the individual computed mode frequencies, but we cannot expect them to be as accurate as the observed mode frequencies. Before presenting the results, we must think about the definition of a "good fit" between the model and observed frequencies. As shown by Deupree (2011b), the inclusion of a sizeable number of basis functions provides so many modes that it is difficult not to match an observed frequency with a computed one, at least if a sufficient range of longitudinal modes is allowed. This is clearly unsatisfying, so we must provide some other criteria to define a satisfactory solution. While these may not be strictly speaking correct in all cases, they are not unreasonable.

One way in which a criterion for a mode to have a reasonable prospect of being observable is to attempt to trace a given mode through progressively more slowly rotating models back to a specific mode of given n,ℓ, and m in the nonrotating model (e.g., Lignières, et al. 2006), with the expectation that low ℓ and m values will correspond to observable modes. Besides being difficult once the rotation rate is sufficiently large, the latitudinal profile of the perturbation can change considerably from the nonrotating model to one that is rotating as rapidly as α Oph. Thus, the connection between ℓ for the mode in the nonrotating model and the latitudinal profile in the rapidly rotating model may not be as tight as necessary for the process to successfully identify observable modes. An alternative, which we adopt here for the highest amplitude modes, is to examine whether the latitudinal variation of the radial perturbation is consistent with being an observable mode at reasonable amplitude. The difficulty in examining the longitudinal variation is that, while we believe the oscillation frequencies are well determined, the corresponding eigenfunctions are less so. While having a certain latitudinal variation may be helpful for the highest amplitude modes, the possible latitudinal cancellation may be considerably more gray for these modes than for pure Legendre polynomial modes, and we do not feel that this approach can be applied for the lower amplitude modes.

Here we shall be concerned only with p modes, rather arbitrarily limiting these to modes above 15 cycles $d^{-1}$. This provides 35 observed modes out of the original 55 observed by MOST. One reason for doing this is that we have assumed that the rotation is uniform, something probably not true in the core and which could affect the g mode frequencies. The first assumption we make about the nature of the pulsation is that the highest amplitude modes are axisymmetric modes which are also symmetric about the equator. This latter assumption is based on the fact that α Oph is observed essentially equator on so that non equatorially symmetric (or odd parity) perturbations would largely cancel out. The limitation to axisymmetric modes is less defensible because it is incomplete. It assumes that modes with no longitudinal modes will have less cancellation than modes with longitudinal modes, but we also need to show that the surface latitudinal variation produces only limited cancellation. The quality of the latitudinal surface variation of the eigenfunction can be assessed by comparing the latitudinal profiles of the same mode when computed with six or eight basis functions, but this is still not a guarantee that they are reliable. Table 2 provides a list of the nine highest amplitude modes in our selected set detected by MOST (Monnier et al. 2010), along with our computed frequency, and the difference between the computed and observed frequencies in units of the 0.017 cycles $d^{-1}$ quoted observational uncertainty. The table shows that all of the four highest amplitude modes are matched in frequency (almost) within the observational error. Three of the next five highest amplitude modes match within the observational error as well. Only seven more computed axisymmetric modes match the next 18 highest amplitude modes (amplitudes all $\geq 0.1$ mmag), and no computed axisymmetric modes match the frequencies of the remaining eight lowest amplitude modes.

We now turn to the latitudinal surface variation of the seven axisymmetric modes which match the observed frequencies. Four of these, modes 2, 3, 4, and 8 present no difficulties. The latitudinal variation of these modes is shown in Figure 3. The relative amplitude near the equator is large in all cases. Mode 2 has two nodes, but they are both at high latitude and the amplitude of the oscillation between the two nodes is small, limiting the cancellation effects. Modes 3 and 8 have only one node, and mode 4 has none. One would expect these to have relatively large observed amplitudes compared to modes with many nodes, and they would appear to do that. The surface latitudinal variation of these modes is not appreciably changed by decreasing the number of basis functions to six. The surface latitudinal variation for the axisymmetric approximations to modes 1, 6, and 9 present more cancellation, shown in Figure 4. The frequencies of modes 1, 5, and 6 can be matched by modes with latitudinal surface variations similar to those in Figure 3 with values of m of -3, 1, and -1, respectively. We have found no mode at all with a frequency close to that for mode 7, while mode 9 can be matched by a mode with m = -3, although the horizontal variation is not as obviously favorable as those in Figure 3. This perhaps should not be surprising given its lower amplitude. Noting that this analysis makes no allowance for the theoretical uncertainty of the mode frequency, we regard such agreement for the seven of the nine high amplitude modes as acceptable.

With the success in matching a number of the high amplitude modes which were both axisymmetric and equatorially symmetric (or even parity), we next examined the possibility that the lowest amplitude modes might also be odd parity axisymmetric modes. Here we found that two of the four lowest amplitude observed mode frequencies could be matched by these modes.

Next we extended the search to include nonaxisymmetric even parity modes having $|m| \leq 2$ to examine the remaining p modes. The quantity m, half of the number of longitudinal nodes, remains a valid quantum number because of axial symmetry in our 2D rotating models. This does produce a large number of modes, but not as many as computed by Deupree (2011b) who allowed up to $|m| = 4$. Finally, we computed nonaxisymmetric odd parity modes with $|m| \leq 2$ to attempt to match the other lowest amplitude modes which were not matched with m = 0.

The frequency match results for all 35 observed modes are shown in order of decreasing amplitude in Table 3. The table includes a number to identify the mode, the observed frequency and amplitude, the computed frequency, the magnitude of the frequency difference in units of the observed frequency uncertainty, and the m value of the computed mode which generated the matching frequency. The designation "odd" is used to indicate odd parity modes. In a few cases more than one computed mode, each with a different value of m, may match the observed mode to within the observational error. Our approach has been to select the mode with the lowest magnitude of m in this case, even if other modes may have had a smaller difference between the observed and computed frequencies. For modes whose smallest difference is outside the observational uncertainty, we selected the mode with the smallest difference. The table shows that 2**3** of the 35 modes were matched to within 1.1 times the observational uncertainty. A further five modes (modes 5, 6, 12, 14, and 15) were matched to within 1.5 times the observational frequency uncertainty, and four more (modes 10, 18, 28, and 31) were matched if the range is extended to twice the observational frequency uncertainty. While subjective, we believe that all of these could be considered reasonable fits given the computed mode uncertainty as well as the observational uncertainty. This leaves only three observed modes (modes 7, 16, and 35) which are not reasonably well matched (i.e., frequency difference outside twice the observational uncertainty) by computed modes. Of these the most worrisome is mode 7, which has fairly large amplitude.

Even these three modes can be matched by expanding the computed mode criteria. Mode 7 can be matched to within the observational error by an odd parity mode with m = 2, even though its high amplitude makes this fairly unattractive. Mode 16 can also be matched to within the observational error by an odd parity mode with m = -2, and its amplitude is sufficiently low that this may not be unreasonable. The lowest amplitude mode, mode 35, could not be matched by any mode with $|m| \leq 2$, but it can be matched to within the observational error with an even parity m = 3 mode. Given its low amplitude, this may not be absurd. Thus, it is not impossible to match the mode frequencies reasonably successfully, but given the density of modes, this does not prove that this model is the correct model of α Oph.

We present the echelle diagram for all of these computed modes (except the odd parity mode matching mode 7) in Figure 5. Echelle diagrams are usually used to show patterns in the frequency spectrum, and of course here there is no simple pattern. However, the diagram also gives a representation of the density of computed modes at various frequencies and how close those frequencies are to the observed frequencies for this model. The figure includes all computed modes for |m| ≤ 2, all the **odd parity** computed modes listed in Table 3, and the m = 3 computed mode matching the lowest amplitude observed mode. The large separation of 3.78 cycles $d^{-1}$ was found by computing axisymmetric modes to frequencies of about 60 cycles $d^{-1}$.

All of the computed modes in Figure 5 have a comparatively small number of radial nodes. While there is some variation, the axisymmetric modes in the lowest part of Figure 5 have only three or four radial nodes (it should be noted that the number of radial nodes for a model rotating this rapidly may be a function of latitude, e.g., Clement 1998). Those at the highest frequencies of Figure 5 have ten or eleven radial nodes. The frequency shifts for nonaxisymmetric modes are large and nonuniform (Suárez, et al. 2010, Deupree & Beslin 2010) for models rotating this rapidly so that any given horizontal line in Figure 5 may contain modes with values of n differing by two or three.

The risky nature of using the latitudinal variation to correlate with amplitude may be seen by comparing modes 4 and 23. The latitudinal variation is the same for both (see Figure 3), so they clearly have the same "ℓ". Mode 4 has six radial nodes in the pressure perturbation in the equatorial direction and four nodes along the polar axis, and mode 23 has one more node at both the equator and pole directions. Thinking of these two modes as having "n, $\ell_{eff}$, and m" quantum numbers, we see they differ only by Δn = 1, while the observed amplitude ratio is almost a factor of three.

## 4. DISCUSSION

We have shown that we could match both the observed spectral energy distribution and the p mode oscillation frequencies of α Oph reasonably well with a uniformly rotating 2.25 $M_\odot$ model having a surface equatorial rotational velocity of 236 km $s^{-1}$ and a central hydrogen abundance of 0.25. While not perfect, the agreement between the observed and computed SED and between the higher amplitude observed modes and computed modes with reasonable latitudinal variation provides perhaps the strongest evidence that this model provides a reasonable approximation to the star. The most glaring exception is the seventh highest amplitude mode at 23.63 cycles $d^{-1}$, which cannot be matched to within four times the observed error with the most reasonable mode constraints, but can be matched within the observed error if we allow the mode to be equatorially asymmetric. Given that α Oph is seen nearly equator on and the amplitude of the mode is relatively large, this assumption seems questionable.

Based on a comparison of these frequency results with those of Deupree (2011b), it is clear that being able to match the observed SED in the both the visual and ultraviolet plays a crucial role in constraining the model. We would argue that models of rapidly rotating δ Scuti stars have a reasonable chance of success in matching observed p mode pulsation frequencies if both interferometry and a SED including both the ultraviolet and the visual region are available. The interferometry is required to identify both the shape of the model surface and the inclination, while the SED imposes reasonable constraints on the surface temperature distribution and equatorial radius. We note that we do not get the shape of the model surface for stars being seen close to pole on. Without these three sets of observational information, the parameter space may simply be too large to bring about a realistic prospect of success.

The large number of computed modes shown in Figure 3 arises from the sizeable number of basis functions needed to obtain reasonably accurate computed frequencies. This large number brings into question what good agreement with the observed oscillation frequencies actually entails. It would be desirable to utilize the horizontal variation of the displacements at the model surface to argue which modes would have a higher probability of being observed, but the computed perturbations are not nearly as accurately determined as the oscillation frequencies are and may show significant variation as the number of basis functions increases (Deupree 2011b). Including yet more basis functions will increase the accuracy of the surface perturbations for some of the modes, but at the expense of adding yet more modes. It should be noted that the deviation in latitudinal variation from that of a single Legendre polynomial may be considerable for sufficiently rapidly rotating stars. These deviations can be such that simple arguments about the number of latitudinal nodes, or the effective $\ell$ value, may be inadequate as a metric for observability. It is not clear how this issue can be resolved.

There remain issues which have not been adequately addressed and which we may not be able to solve. The most important of these is perhaps the internal angular momentum distribution. It is clear that this model produced the observed ratio of polar to equatorial radius with uniform rotation, leading to the expectation that the star may not be too far from uniform rotation, at least near the surface. This assumes that the surface is an equipotential, something that, while not unreasonable, remains an assumption. One would expect for a star as evolved as α Oph that a more reasonable angular momentum distribution would not correspond to a conservative rotation law because at least part of the core would be rotating more quickly than the surface, even at the pole, in which case no equipotential can be defined. It is hoped, but certainly not proven, that a study of the low frequency modes might provide constraints on both the internal angular momentum distribution and the composition profiles.

Despite these concerns, we do believe that this 2D model for α Oph does provide us with our best estimate to date concerning what we can and cannot expect to be able to do with oscillation frequencies for at least moderately rapidly rotating δ Scuti stars.

This work is supported in part by an ACEnet Postdoctoral Fellowship to FP and an National Sciences and Engineering Research Council of Canada (NSERC), Discovery grant to CIS. High performance computing (HPC) facilities for many of these calculations through ACEnet, the HPC consortium for universities in Atlantic Canada and through a Canada Foundation for Innovation (CFI) grant to RGD. ACnet is supported by CFI, NSERC, the Atlantic Canada Opportunities Agency, the Nova Scotia Research Innovation Trust, and the research and innovation funding agencies of the provinces of New Brunswick and Newfoundland and Labrador.

Table 1

Comparison of Observed and Computed Properties

| Property | Observed Value | Computed (or assumed) value |
|---|---|---|
| $V_0$ | 2.08[1,2,3] | 2.074 |
| $(B-V)_0$ | 0.15[1,2,3] | 0.158 |
| $(U-B)_0$ | 0.10[1,2,3] | 0.124 |
| $R_{pole} / R_{eq}$ | 0.836[4] | 0.838 assumed |
| Distance (pc) | 14.68[5] | 14.6 |
| $V_{eq}$ (km $s^{-1}$) | 210.- 240.[6,7,8,9] | 236. |
| Inclination (degrees) | 87.5[4] | 87.5 assumed |
| Mass ($M_\odot$) | 2.4[10] | 2.25 (assumed) |

[1]Johnson & Harris (1954)

[2]Johnson & Knuckles (1957)

[3]Johnson, et al. (1966)

[4]Monnier, et al. (2010)

[5]Gatewood (2005)

[6]Bernacca & Perinotto (1970)

[7]Uesugi & Fukuda (1970)

[8]Abt & Morrell (1995)

[9]Royer, et al. (2002)

[10]Hinkley, et al. (2011)

Table 2

Comparison of Observed and Axisymmetric Computed Frequencies for Highest Amplitude Modes

| Observed Frequency (cycles day$^{-1}$) | Half-Amplitude (millimag) | Computed Frequency (cycles day$^{-1}$) | Difference / Observational Uncertainty |
|---|---|---|---|
| 18.668 | .655 | 18.6724 | 0.259 |
| 16.174 | .411 | 16.1700 | 0.235 |
| 16.124 | .349 | 16.1202 | 0.224 |
| 22.205 | .299 | 22.2226 | 1.037 |
| 21.713 | .282 | 21.7617 | 2.865 |
| 25.416 | .272 | 25.4323 | 0.956 |
| 23.631 | .223 | 23.7062 | 4.422 |
| 20.512 | .210 | 20.5014 | 0.624 |
| 29.304 | .172 | 29.3168 | 0.751 |

Table 3

Comparison of Computed and Observed Mode Frequencies for All Modes

| Mode ID | Observed Mode Frequency (cycles day$^{-1}$) | Half-Amplitude (millimag) | Computed Mode Frequency (cycles day$^{-1}$) | $\lvert\Delta\omega\rvert/\sigma_{obs}$ | Mode m Value |
|---|---|---|---|---|---|
| 1 | 18.668 | 0.655 | 18.6670 | .059 | m = -3 |
| 2 | 16.174 | 0.411 | 16.1700 | .235 | m = 0 |
| 3 | 16.124 | 0.349 | 16.1202 | .224 | m = 0 |
| 4 | 22.205 | 0.299 | 22.2226 | 1.037 | m = 0 |
| 5 | 21.713 | 0.282 | 21.7355 | 1.325 | m = 1 |
| 6 | 25.416 | 0.272 | 25.4428 | 1.576 | m = -1 |
| 7 | 23.631 | 0.223 | 23.7008 | 4.103 | m = -2 |
| 8 | 20.512 | 0.210 | 20.5014 | .624 | m = 0 |
| 9 | 29.304 | 0.172 | 29.3103 | .370 | m = -3 |
| 10 | 43.292 | 0.157 | 43.3239 | 1.878 | m = -1 |
| 11 | 20.286 | 0.151 | 20.2738 | .720 | m = 2 |
| 12 | 35.718 | 0.149 | 35.6955 | 1.323 | m = 0 |
| 13 | 39.597 | 0.149 | 39.5994 | .141 | m = 1 |
| 14 | 22.480 | 0.146 | 22.4569 | 1.362 | m = 1 |
| 15 | 25.166 | 0.144 | 25.1881 | 1.299 | m = 1 |
| 16 | 24.582 | 0.136 | 24.6427 | 3.572 | m = 2 |
| 17 | 20.420 | 0.120 | 20.4281 | .475 | m = 2 |
| 18 | 22.155 | 0.117 | 22.1835 | 1.675 | m = -2 |
| 19 | 17.183 | 0.115 | 17.1823 | .040 | m = -2 |
| 20 | 25.250 | 0.114 | 25.2339 | .947 | m = 1 |
| 21 | 35.877 | 0.113 | 35.8857 | .511 | m = 1 |
| 22 | 27.001 | 0.112 | 27.0063 | .312 | m = -2 |
| 23 | 25.633 | 0.111 | 25.6277 | .311 | m = 0 |
| 24 | 19.252 | 0.111 | 19.2335 | 1.088 | m = 0 |
| 25 | 18.818 | 0.109 | 18.8166 | .080 | m = -1 |
| 26 | 19.936 | 0.109 | 19.9322 | .225 | m = 0 |
| 27 | 29.120 | 0.105 | 29.1296 | .565 | m = 0 |
| 28 | 20.228 | 0.093 | 20.1944 | 1.976 | m = -1 (odd) |
| 29 | 23.807 | 0.092 | 23.8031 | .230 | m = -1 (odd) |
| 30 | 18.209 | 0.091 | 18.2238 | .872 | m = -2 |
| 31 | 31.130 | 0.089 | 31.1582 | 1.659 | m = 2 (odd) |
| 32 | 30.980 | 0.072 | 30.9649 | .886 | m = 0 (odd) |
| 33 | 32.949 | 0.060 | 32.9469 | .123 | m = 1 |
| 34 | 34.392 | 0.049 | 34.4093 | 1.018 | m = 0 (odd) |
| 35 | 48.347 | 0.036 | 48.2973 | 2.925 | m = 1 (odd) |

Figure Captions

Fig. 1a. – Spectral energy distribution for a spherical stellar model with $T_{eff}$ = 7875K and log $g_{eff}$ = 3.67 (solid curve) compared to that observed (dashed curves and circles for selected points) for α Oph. The agreement in the visual region in essence determines the effective temperature. Note the disagreement just to the blue of the Balmer jump.

Fig. 1b. – Spectral energy distribution in the ultraviolet for the spherical model in Fig. 1a (solid curve) and the observations (dashed curves). While the general agreement in the ultraviolet is fair, over much of the interval the data fluxes are higher than the computed model fluxes.

Fig. 2a. – Spectral energy distribution seen at an inclination of 87.5$^o$ for the rapidly rotating model discussed in the text (solid curve) and the observed SED (dashed curves and circles for selected points) for α Oph. This model also provides a good fit to the visual SED although the equatorial effective temperature is about 140K lower than the effective temperature of the spherical model in Figure 1.

Fig. 2b. – The ultraviolet spectral energy distribution seen at an inclination of 87.5$^o$ for the rapidly rotating model discussed in the text (solid curve) and the observations for α Oph (dashed curves). In comparison with Fig. 1b, note that the rotating model SED fits the observations in the ultraviolet and to the blue of the Balmer jump appreciably better than does the spherical model SED.

Fig. 3. – The surface latitudinal variation of the radial perturbation for the axisymmetric eigenfunctions which match the oscillation frequencies for modes 2, 3, 4, and 8 as given in Table 2. The linear perturbations are scaled to be unity at 80$^o$ colatitude. The dashed lines indicate “0” for each of the four modes. The small number of latitudinal nodes for each mode should allow them to be observable at a reasonable amplitude.

Fig. 4. – The surface latitudinal variation of the radial perturbation for the axisymmetric eigenfunctions which match the oscillation frequencies for modes 1, 6, and 9 as given in Table 2. The linear perturbations are scaled to be unity at 10$^o$ colatitude. The number of nodes and their placement suggest that it might be difficult to observe these modes at the amplitudes the observed modes display.

Fig. 5. – Echelle diagram for the computed and observed frequencies of α Oph. The data are denoted by circles and are vertically offset slightly for clarity. Squares are axisymmetric modes, diamonds represent m = 1 modes, inverted triangles m = 2 modes, a triangle the sole m = 3 mode to match the highest

observed frequency mode, triangles pointing to the right m = -1, triangles pointing to the left m = -2, and asterisks for the equatorially asymmetric modes given in Table 3.

Figure 1a

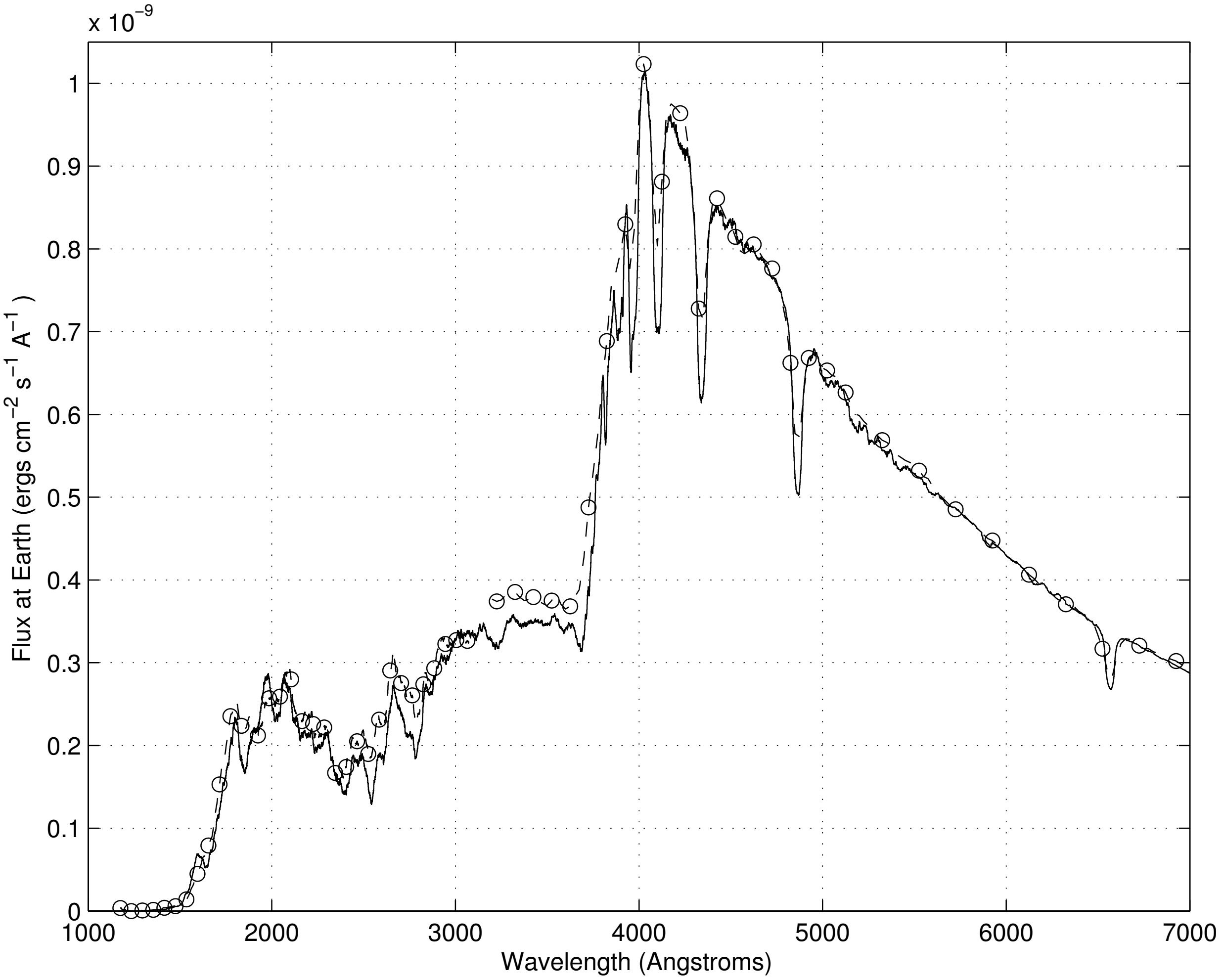

x 10^-9
1
0.9
0.8
0.7
0.6
0.5
0.4
0.3
0.2
0.1
0
Flux at Earth (ergs cm^-2 s^-1 A^-1 )
1000
2000
3000
4000
5000
6000
7000
Wavelength (Angstroms)

Figure 1b

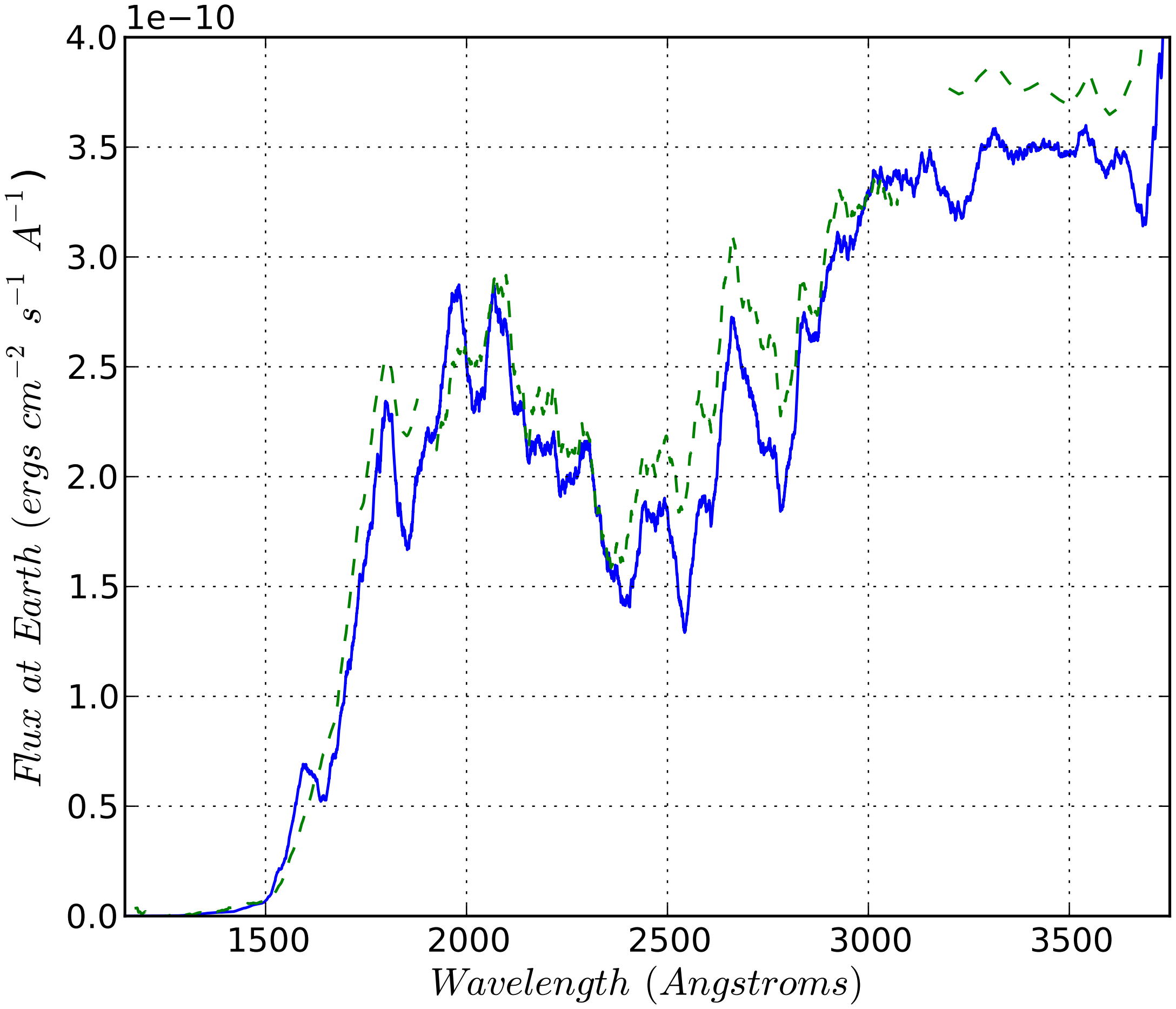

Figure 2a

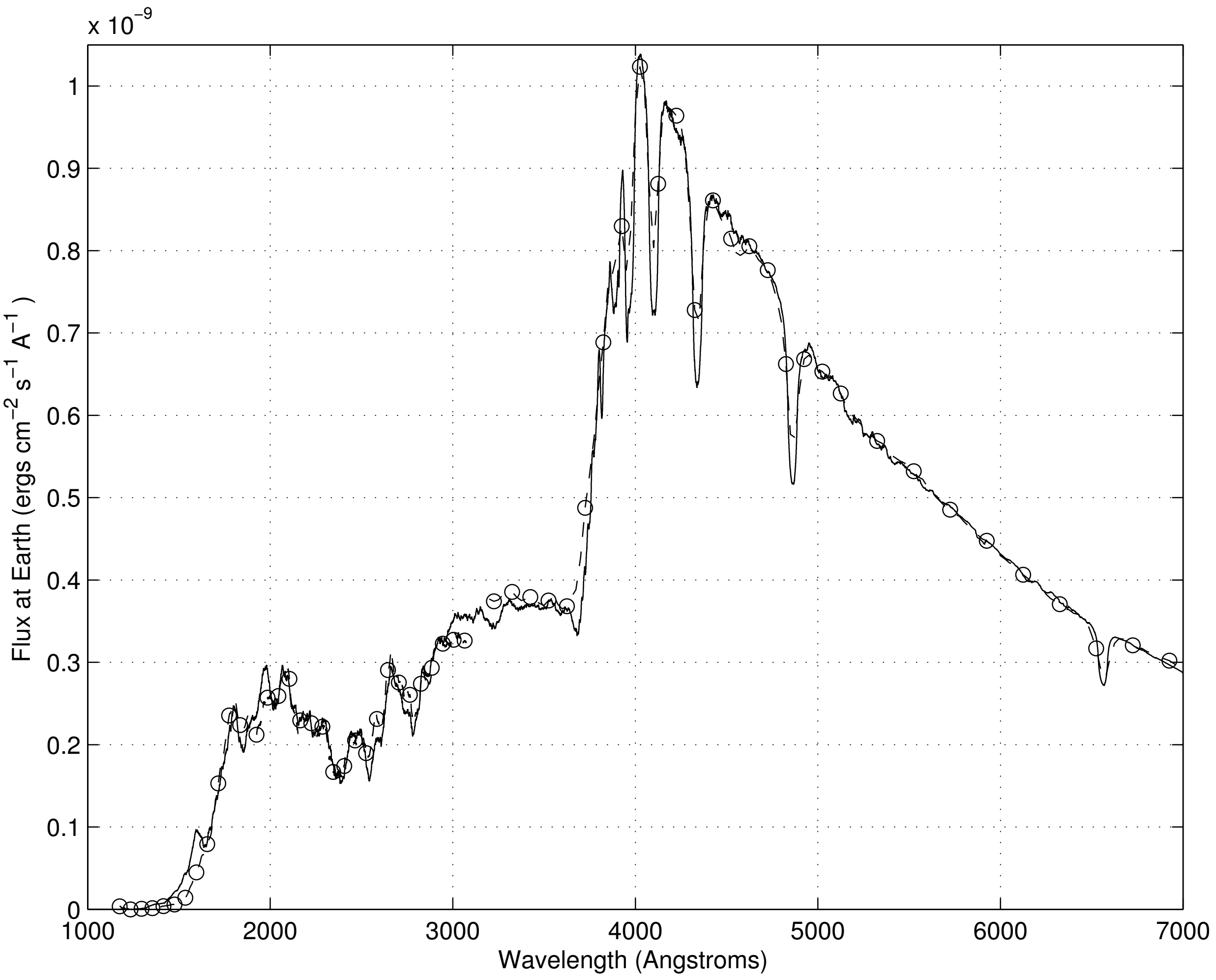

x 10^-9
1
0.9
0.8
0.7
0.6
0.5
0.4
0.3
0.2
0.1
0
Flux at Earth (ergs cm^-2 s^-1 A^-1 )
1000
2000
3000
4000
5000
6000
7000
Wavelength (Angstroms)

Figure 2b

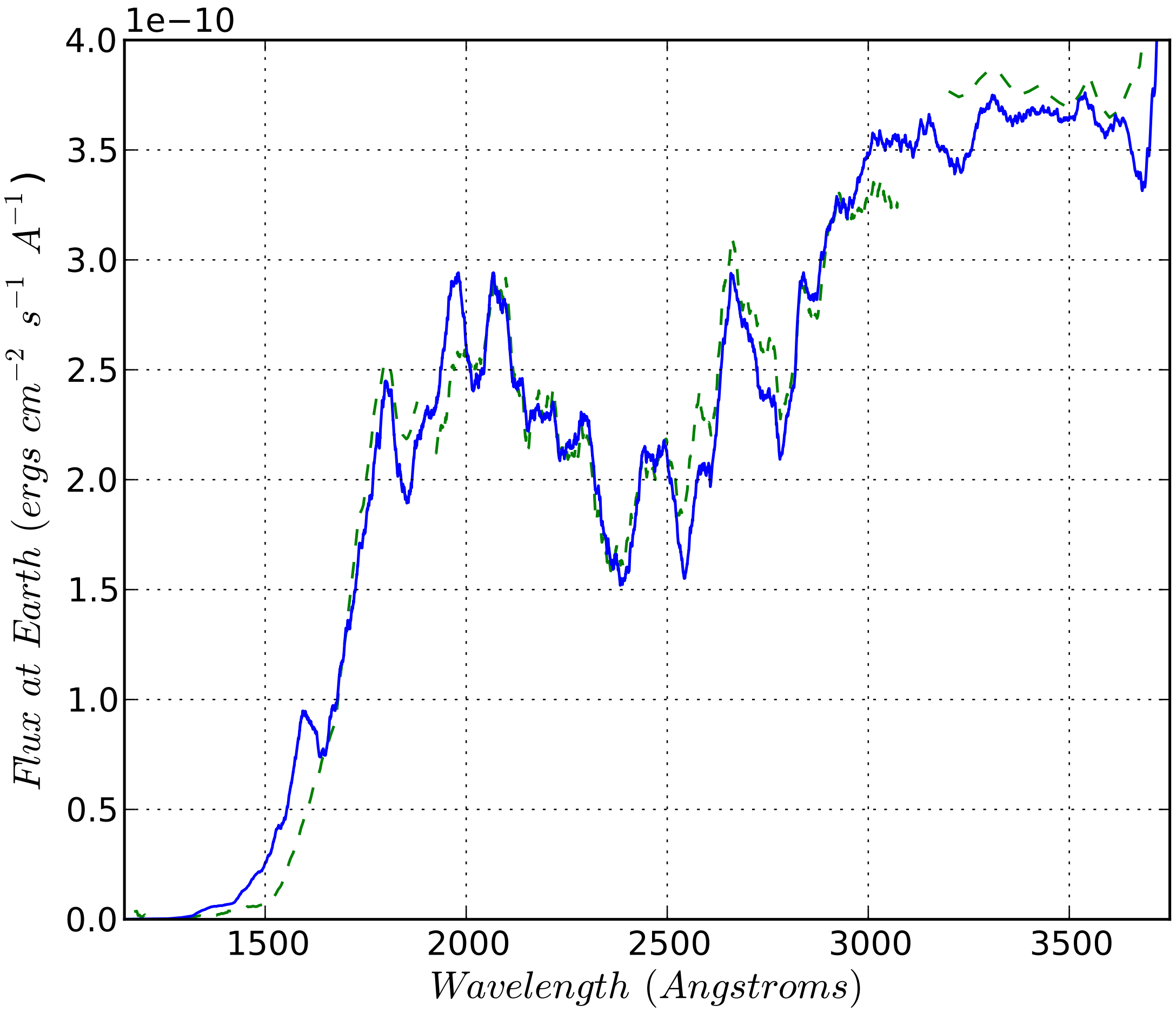

Figure 3

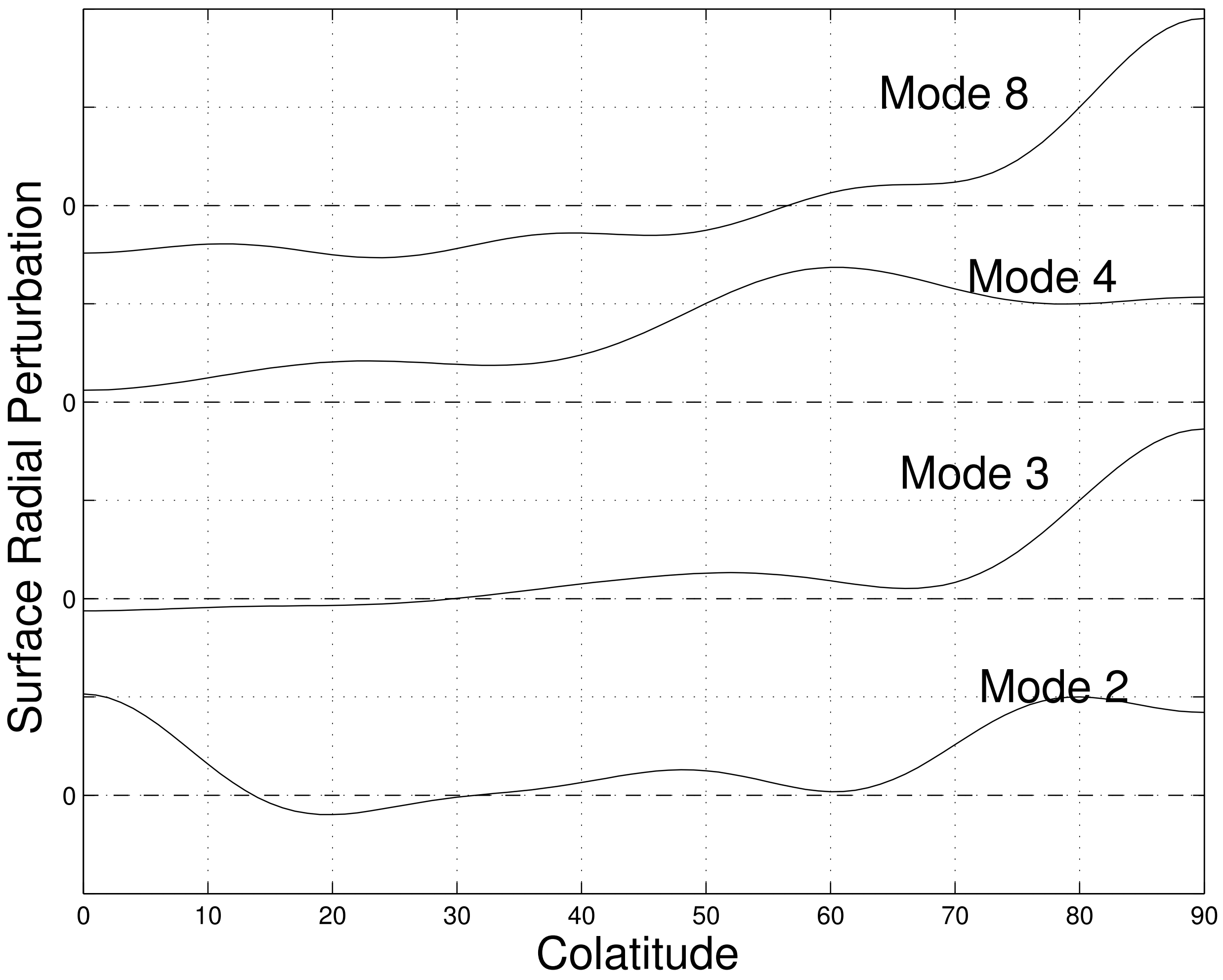

Figure 4

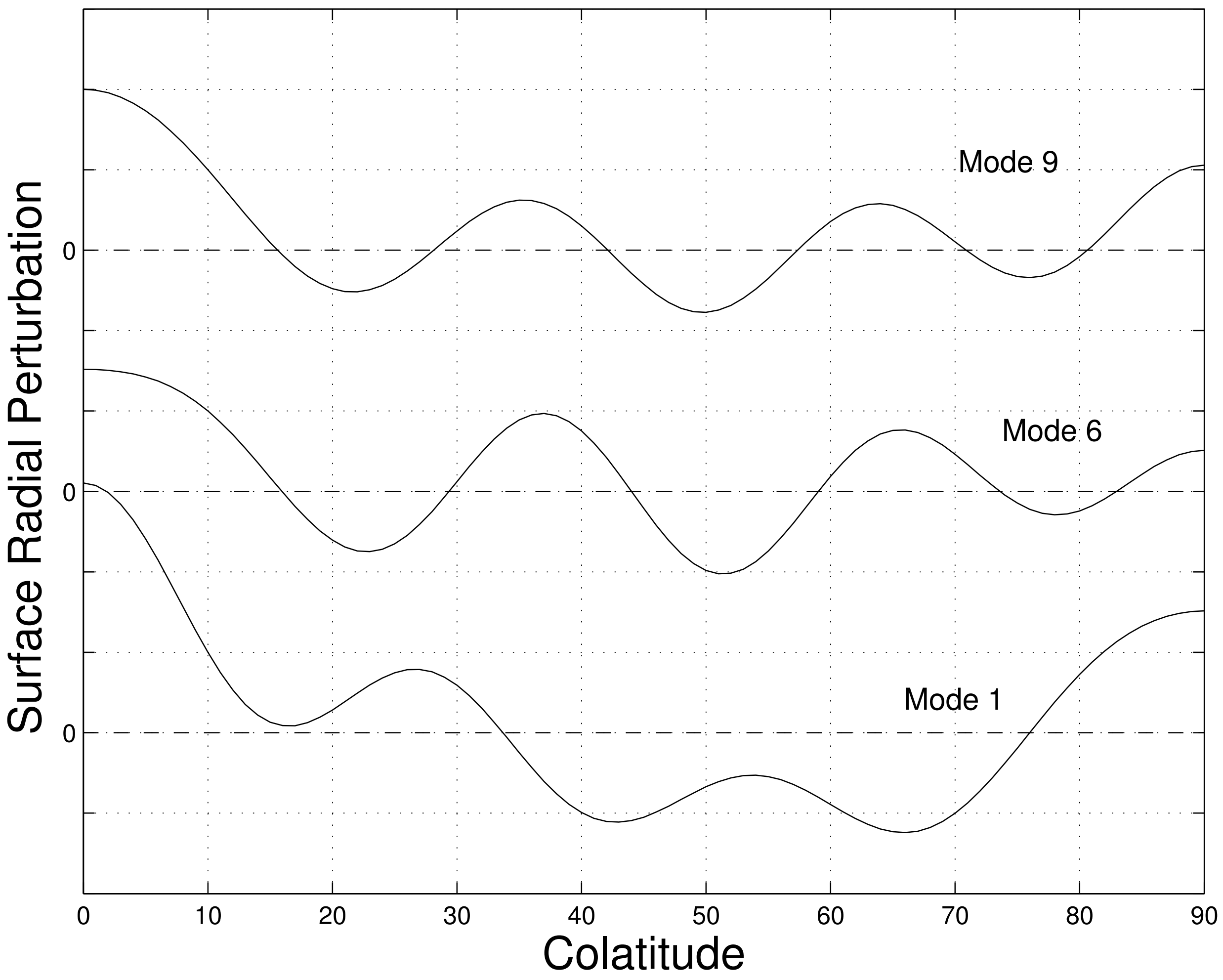

Figure 5

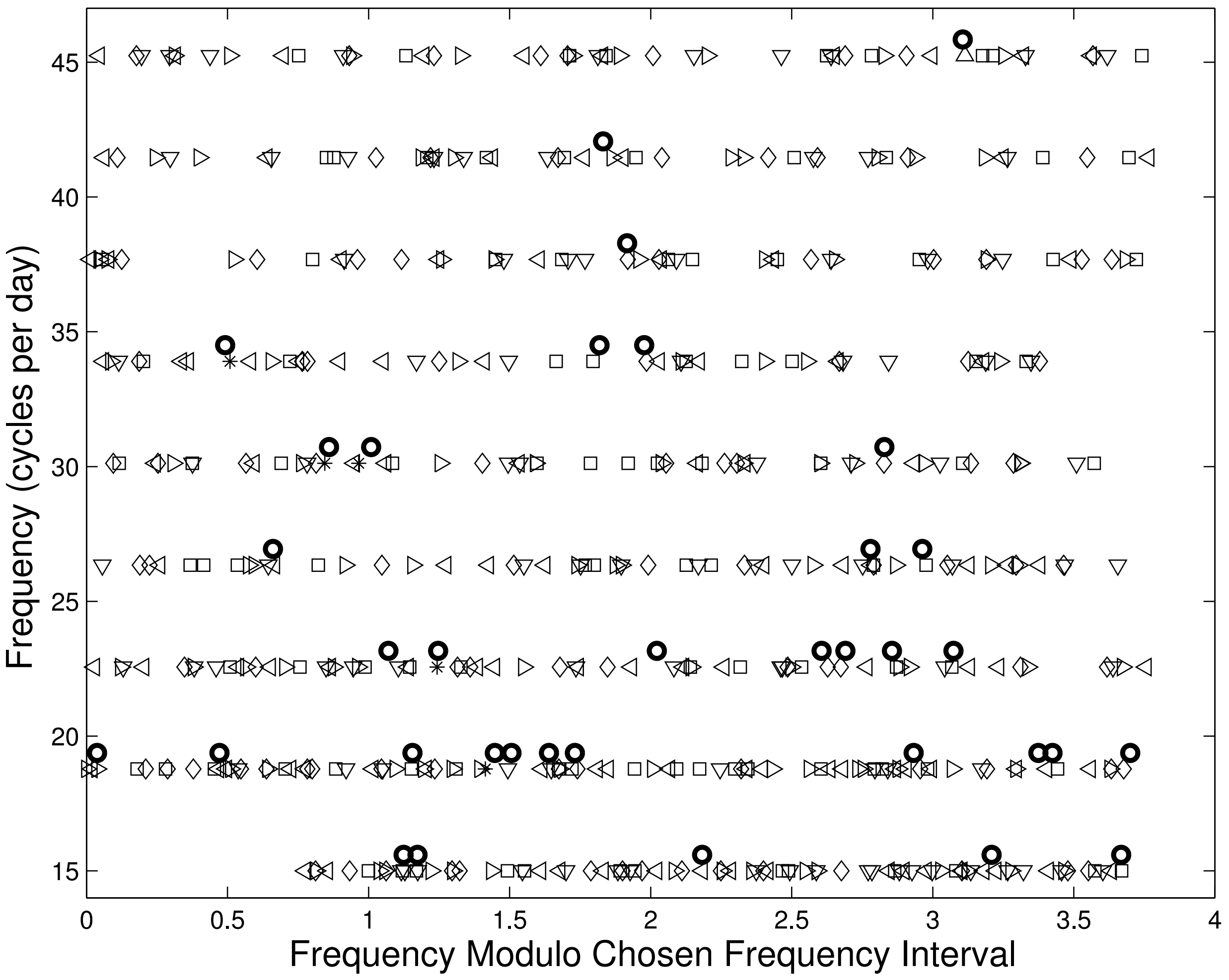